\documentclass[aps,pra,twocolumn,superscriptaddress,showpacs,longbibliography]{revtex4-1}

\usepackage{graphicx}
\usepackage{amsmath}
\usepackage{color}

\usepackage[bookmarks=true,colorlinks=true,urlcolor=blue,linkcolor=blue,citecolor=blue,breaklinks]{hyperref}

\usepackage{cleveref}
\usepackage{amssymb}
\usepackage{booktabs}

\usepackage{float}

\begin{document}

\sloppy






\title{Broken adiabaticity induced by Lifshitz transition in MoS$_2$ and WS$_2$ single layers}


\newcommand*{\DIPC}[0]{{
Donostia International Physics Center (DIPC),
Paseo Manuel de Lardizabal 4, 20018 Donostia-San Sebasti\'an, Spain}}

\newcommand*{\IFS}[0]{{
Center of Excellence for Advanced Materials and Sensing Devices, Institute of Physics, Bijeni\v{c}ka 46,
10000 Zagreb, Croatia}}

\author{Dino Novko}
\email{dino.novko@gmail.com}
\affiliation{\IFS}
\affiliation{\DIPC}


\begin{abstract}

The breakdown of the adiabatic Born-Oppenheimer approximation is striking dynamical phenomenon, however, it occurs only in a handful of layered materials. Here, I show that adiabaticity breaks down in doped single-layer transition metal dichalcogenides in a quite intriguing manner. Namely, significant nonadiabatic coupling, which acts on frequencies of the Raman-active modes, is prompted by a Lifshitz transition due to depopulation and population of multiple valence and conduction valleys, respectively. The outset of the latter event is shown to be dictated by the interplay of highly non-local electron-electron interaction and spin-orbit coupling. In addition, intense electron-hole pair scatterings due to electron-phonon coupling are inducing phonon linewidth modifications as a function of doping. Comprehending these intricate dynamical effects turns out to be a key for mastering characterization of electron doping in two-dimensional nano-devices by means of Raman spectroscopy.




%
%

\end{abstract}


\maketitle


The Born-Oppenheimer approximation, which assumes that electrons adiabatically follow the motion of lattice degrees of freedom, is a fundamental starting point in quantum mechanics. Recent studies have, however, shown that striking nonadiabatic (NA) effects visible in vibrational Raman spectra can be found in metallic layered materials, such as graphene\,\cite{bib:lazzeri06,bib:pisana07,bib:ferrante18}, graphite intercalation compounds\,\cite{bib:saitta08}, and magnesium diboride\,\cite{bib:quilty02,bib:saitta08}, as well as in doped semiconductors, such as boron-doped diamond\,\cite{bib:caruso17}. The strength of this dynamical electron-phonon coupling (EPC) can, in fact, be modified as a function of carrier concentration, which makes the Raman spectroscopy a quite powerful tool for characterization of doped two-dimensional and layered materials\,\cite{bib:das08}. However, to make this characterization complete, the precise understanding of the microscopic processes underlying the NA modifications of phonons is needed\,\cite{bib:lazzeri06,bib:saitta08,bib:giustino17,bib:novko18}. Such renormalization of phonons is actually considered quite rare, and apart from these few cases it is not clear in what other two-dimensional materials should adiabaticity break down.

Contrary to the aforesaid examples, the vibrational spectra of the single-layer transition metal dichalcogenides (TMDs) under bias voltage is believed to be adiabatic\,\cite{bib:zhang15,bib:chakraborty12,bib:ponomarev19}. In other words, the gate-induced phonon renormalization can be qualitatively described by means of adiabatic density functional theory (DFT) calculations\,\cite{bib:chakraborty12}. Despite this common belief, there are few worthwhile indications that point to the importance of the NA effects in doped TMDs\,\cite{bib:ponomarev19,bib:zheng17,bib:zheng18,bib:reichardt19}. For instance, recent study on carrier-induced phonon modifications in atomically-thin TMDs had shown that the adiabatic DFT results are largely overestimating the corresponding frequency shifts of the Raman-active modes, but NA corrections were not taken into account\,\cite{bib:ponomarev19}. Moreover, the theoretical considerations have predicted the importance of the NA coupling in charge transfer dynamics of TMDs heterostructures\,\cite{bib:zheng17,bib:zheng18} as well as in exciton-mediated Raman scattering of MoS$_2$\,\cite{bib:reichardt19}. Additionally, the effect beyond adiabatic approximation, namely, the phonon-mediated superconductive state was shown to exist in single- and few-layer MoS$_2$ when multiple conduction valleys are partially occupied\,\cite{bib:ge13,bib:costanzo16,bib:piatti18}. Considering the plethora of exceptional charge-induced optical properties\,\cite{bib:mak12,bib:jo14,bib:wang15,bib:wang17} and good carrier mobility in doped single-layer TMDs\,\cite{bib:radisavljevic11}, as well as the pivotal role of the EPC in these features, it is of paramount importance to decipher the role of the NA effects and the EPC in the corresponding Raman spectra.

Here, I show that the vibrational spectra of single-layer MoS$_2$ and WS$_2$, as prototypical examples of semiconducting TMDs, doped with electrons and holes are indeed governed by strong NA effects. Namely, when more than one valley of either valence or conduction electronic bands are simultaneously partially occupied or empty a Lifshitz transition occurs (i.e., an abrupt change of the Fermi surface)\,\cite{bib:piatti18} which then activates the significant dynamical part of the EPC. The latter is manifested as a considerable redshift of the $A_{1g}$ optical phonon frequency along with the large corresponding NA correction. In addition, there are other intriguing coupling mechanisms ruling the NA renormalization of phonons in TMDs. Firstly, the energy difference between the top (bottom) of the valence (conduction) band valleys, and thus the amount of carriers needed in order to empty (fill) both of them, is highly sensitive to the non-local electron-electron interaction. Furthermore, the spin-orbit coupling (SOC) is important in order to correctly capture the EPC in TMDs. And lastly, a high degree of the electron-hole pair scattering\,\cite{bib:novko18} is present in doped TMDs, which in turn slightly washes out the NA renormalization effects and is responsible for the enlargement of the phonon linewidths as a function of doping. All these processes are essential for recreating the experimental observations and therefore the breakdown of the Born-Oppenheimer approximation in doped TMDs turns out to be even more intricate than in the well-know examples\,\cite{bib:lazzeri06,bib:saitta08,bib:caruso17}.

\section*{Results}

\begin{figure*}[!ht]
\includegraphics[width=0.97\textwidth]{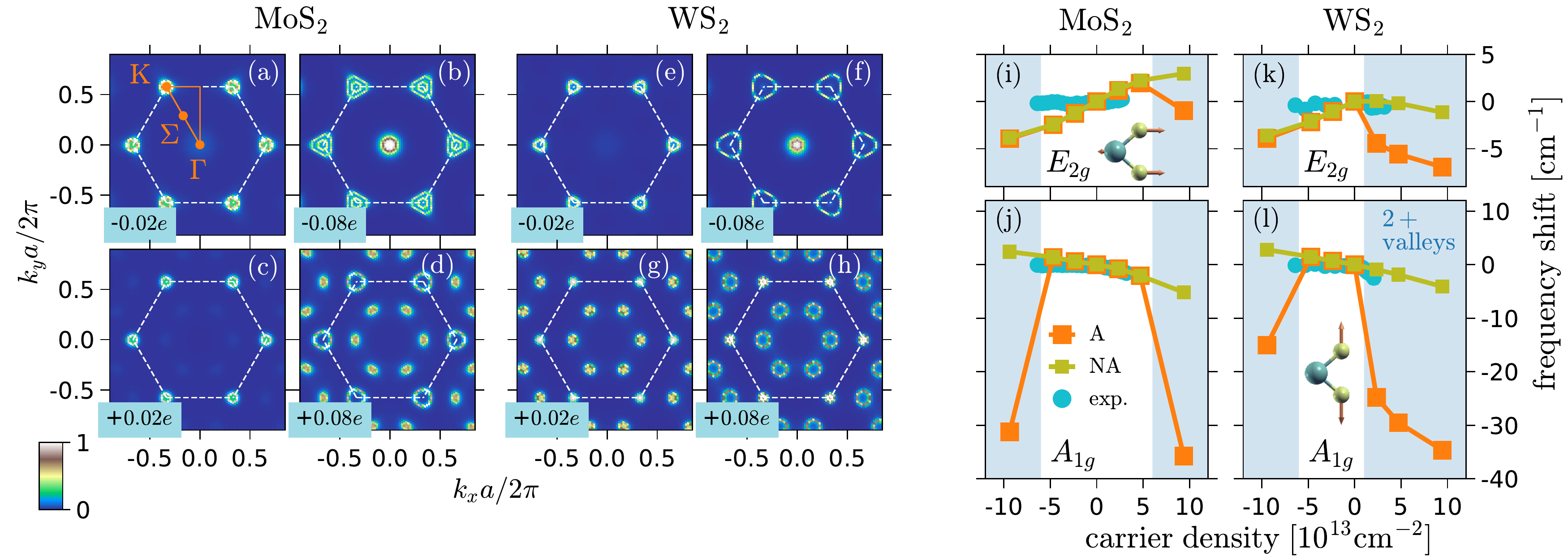}
\caption{\label{fig:fig1}{\bf Fermi surfaces and nonadiabatic phonon frequency shifts.} Color plots of the simulated Fermi surface cuts for different electron ($+e$) and hole ($-e$) dopings in the case of (a)-(d) MoS$_2$ and (e)-(h) WS$_2$. The dopings $\pm0.02e$ and $\pm0.08e$ correspond to carrier densities of $\pm2.34\cdot 10^{13}$\,cm$^{-2}$ and $\pm9.38\cdot 10^{13}$\,cm$^{-2}$, respectively. Orange triangle is the irreducible part of the Brillouin zone and orange dots show the relevant high-symmetry points. Fermi surfaces are shown for each electron momentum along \textit{x} and \textit{y} directions, i.e., $k_x a / 2\pi$ and $k_y a / 2\pi$ (where \textit{a} is unit cell parameter). The frequency shifts of the $E_{2g}$ and $A_{1g}$ phonons as a function of carrier density are shown in (i)-(j) for MoS$_2$ and in (k)-(l) for WS$_2$. The adiabatic (A) and nonadiabatic (NA) results (based in density functional theory) are depicted with orange and green squares, respectively. The experimental results from Ref.\,\cite{bib:ponomarev19} are shown with blue circles. The insets in (i) and (l) display the atomic displacements for the  $E_{2g}$ and $A_{1g}$ phonon modes, respectively. The blue shaded areas in (i)-(l) define the instants when the Fermi level crosses two and more valleys (i.e., the outset of the Lifshitz transition).
}
\end{figure*}

\textbf{Topology of valence and conduction bands.} According to the experimental studies\,\cite{bib:miwa15,bib:bussolotti19,bib:eknapakul14,bib:dendzik15,bib:kastl18}, the electronic structure of the single-layer semiconducting TMDs is characterized with the two valleys in both the conduction and valence bands. For the former, the global minimum is located in the K point of the Brillouin zone, while the remnant minimum is at the $\Sigma$ point, i.e., halfway along the $\Gamma -{\rm K}$ path [see Fig\,\ref{fig:fig1}(a)]. The valence band has band maxima in the $\Gamma$ and K points, where the latter is the global maximum. The corresponding measured energy differences between these minima and maxima are presented in Table\,\ref{tab:table1}. Since the exact carrier density required for the Fermi level to cross the second valleys centered at the $\Sigma$ and $\Gamma$ points depends on this energy difference, it is first essential to capture the proper electronic band structures of single-layer TMDs before studying in detail the doping-induced phonon frequency shifts.


Table\,\ref{tab:table1} shows the results for the energy difference between the top of the K and $\Gamma$ valleys in the valence band $\Delta\varepsilon^{v}_{{\rm K}\Gamma}$ as well as between the bottom of the $\Sigma$ and K valleys in the conduction band $\Delta\varepsilon^{c}_{\Sigma{\rm K}}$ for MoS$_2$ and WS$_2$ and several electron exchange and correlation approximations. It turns out that the standard density functional theory (DFT) calculations with semi-local Perdew, Burke and Ernzerhof (PBE) approximation are not enough to capture the right topology of the valence band in MoS$_2$, since the order of the two valleys is reversed in this case (see also Supplementary Figures 1-3). Therefore, I utilize higher order of approximation for the electronic ground state. Namely, the self-consistent GW0\,\cite{bib:gw0} as well as hybrid functionals PBE0\,\cite{bib:pbe0} and vdW-DF-cx0\,\cite{bib:cx0}. The corresponding results are, on the other hand, much closer to the experiments, which shows how the topology of the valleys in single-layer TMDs, and not just the energy band gap, is highly dictated by the strong quasi-particle physics, i.e., by highly non-local electron-electron interactions\,\cite{bib:kormanyos15}. The SOC as well plays an important role here. Unfortunately, the DFT phonon calculations are numerically unfeasible when performed on top of the GW or the hybrid-functional electronic structure. In order to overcome this hurdle and having in mind that the strain of the unit cell can induce the valley modifications\,\cite{bib:yuan16,bib:ponomarev19,bib:ortenzi18}, the remaining calculations are performed with the PBE functional and with the exact amount of the strain necessary to recreate the correct order and topology of the valleys in MoS$_2$ and WS$_2$ (i.e., as obtained with the vdW-DF-cx0 functional). Further computational details can be found in the Methods section.

The corresponding Fermi surface cuts as a function of electron and hole dopings are shown in Figs.\,\ref{fig:fig1}(a)-(d) and \ref{fig:fig1}(e)-(h) for MoS$_2$ and WS$_2$, respectively. These Fermi surfaces clearly depict the transformation from the one-valley to multi-valley electronic band structure as the electron and hole densities are elevated (see also Supplementary Figure 4), in close agreement with the experiments\,\cite{bib:yuan16,bib:piatti18}. The abrupt appearance of the $\Gamma$ and $\Sigma$ valleys (e.g., for MoS$_2$ when charge density is around $\pm 0.08\,e/$unit cell) represents the standard case of the Lifshitz transition, which is believed to be responsible for the superconductivity in MoS$_2$\,\cite{bib:piatti18}. Here I show that this sudden change in the electron density of states induces the significant dynamical effects in the EPC contributing to the frequency of the Raman-active phonon modes. 

{\bf Nonadiabatic frequency renormalization.} The calculated doping-induced frequency shifts of the $E_{2g}$ and $A_{1g}$ optical phonon modes in MoS$_2$ and WS$_2$ single layers are shown in Figs.\,\ref{fig:fig1}(i)-(l). In particular, I show the results obtained by means of the adiabatic (A) density functional perturbation theory (DFPT) as well as the frequencies corrected with the DFT-based NA method\,\cite{bib:novko18,bib:caruso17,bib:giustino17} (see the Method section and Supplementary Note 1). The corresponding absolute values of the adiabatic frequencies for pristine systems are given in the Supplementary Table\,1. From this the two distinct regimes emerge in the doped single-layer TMDs. Namely, the A regime where only the K point valleys intersect the Fermi level and the NA regime where the Lifshitz transition occurs and sizable dynamical contribution to the EPC is triggered. This intriguing phenomenon is present in both systems and for the both optical phonon modes. However, the NA effects are more pronounced for the $A_{1g}$ phonons, indicating a larger EPC in this case\,\cite{bib:chakraborty12}. For instance, in both TMDs when carrier density is $\sim 10^{14}$\,cm$^{-2}$ the difference between the NA and A frequencies ($\Delta\omega_{\rm NA}$) of the $A_{1g}$ mode is $\sim 30$\,cm$^{-1}$, while only $\sim 5$\,cm$^{-1}$ for the $E_{2g}$ mode.
Note here that the relative NA frequency correction for the $A_{1g}$ mode ($\Delta\omega_{\rm NA}/\omega_{\rm A} \sim 8\%$) is even larger than for the well-studied case of the $E_{2g}$ mode in graphene ($\Delta\omega_{\rm NA}/\omega_{\rm A} \sim 3\%$)\,\cite{bib:lazzeri06} for the similar amount of carrier density.
The potential importance of the NA effects in pristine TMDs was in fact discussed in L. Ortenzi et al.\,\cite{bib:ortenzi18}, where a strong entanglement between the lattice degrees of freedom and the electronic excitations was found by using electronic band structure calculations. However, the NA corrections to the phonon spectrum were not discussed nor calculated.
The results also show how the NA corrections are crucial for achieving a quantitative agreement with the experimental frequency shifts\,\cite{bib:ponomarev19}, particularly for WS$_2$. A more detailed comparison with the experimental results is depicted in the Supplementary Figure 5. Moreover, the exact carrier density windows defining the two regimes differ in MoS$_2$ and WS$_2$, which is due to different topology of the valence and conduction bands in these two TMDs. Consequently, the latter demonstrates how the Raman spectroscopy could be utilized in order to extract the right amount of carrier density required for one- to multi-valley transformations in TMDs and similar semiconductors. Also, since both superconductive state and the breakdown of the adiabaticity occur at the outset of the Lifshitz transition, it is likely that the nonadiabatic channels are open in the superconductivity of the TMDs (Migdal's theorem is not valid in the nonadiabatic regime\,\cite{bib:engelsberg63}), and thus the framework beyond the standard Migdal-Eliashberg theory\,\cite{bib:grimaldi95} might be necessary to explain the corresponding superconductive pairing as well as the discrepancy between the theoretically- and experimentally-obtained transition temperatures $T_c$\,\cite{bib:piatti18}.

Note that the significant doping-induced redshift of the adiabatic phonon frequencies in TMDs was as well obtained in T. Sohier et al.\,\cite{bib:ponomarev19}, where such behaviour was explained in terms of reduced electrostatic screening caused by the population of the multiple valleys (see the Supplementary Figures 6 and 7 for comparison and further discussion). However, it should be emphasized that the NA effects were disregarded in that case and thus the frequency shifts were considered as a pure adiabatic phenomenon. Similar reasoning was as well delivered in the earlier work\,\cite{bib:chakraborty12}. On the other hand, here I show that the significant redshift of the adiabatic $A_{1g}$ phonon frequency is reduced due to strong dynamical effects [e.g., see Fig.\,\ref{fig:fig1}(l)], which in turn improves the agreement with the experiment.

\begin{figure*}[!ht]
\includegraphics[width=0.97\textwidth]{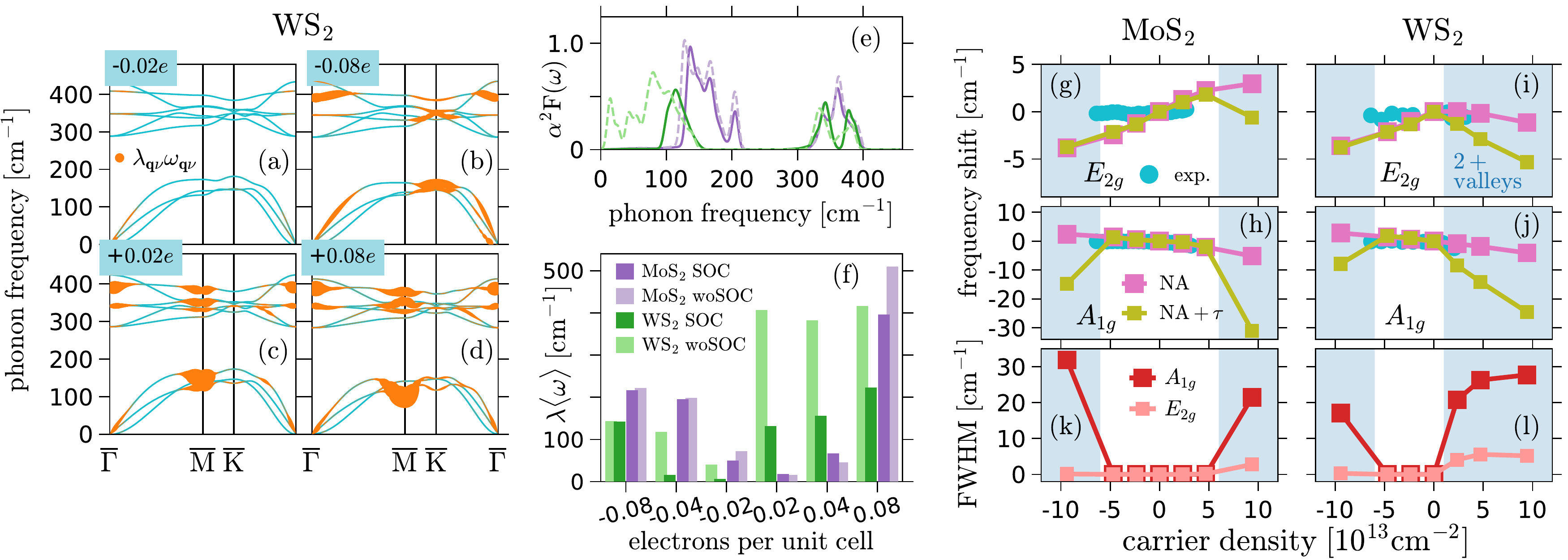}
\caption{\label{fig:fig2}{\bf Role of electron-hole pair scattering processes.} (a)-(d) Phonon band structure for several electron ($+e$) and hole ($-e$) dopings in the case of WS$_2$. In addition, size of the orange circles shows the strength of the momentum- and mode-resolved electron-phonon coupling constants weighted with the corresponding frequencies $\lambda_{\mathbf{q}\nu}\omega_{\mathbf{q}\nu}$. (e) The Eliashberg function $\alpha^2 F(\omega)$ for $0.08\,e/$unit cell and for MoS$_2$ and WS$_2$ when spin-orbit coupling is included (purple and green) or not (light purple and light green). (f) The total amount of the electron-hole pair scattering rate due to electron-phonon coupling for several dopings and for MoS$_2$ (purple) and WS$_2$ (green). The light shades of purple and green are the results without the spin-orbit coupling. The dopings $\pm0.02e$, $\pm0.04e$, and $\pm0.08e$ correspond to carrier densities of $\pm2.34\cdot 10^{13}$\,cm$^{-2}$, $\pm4.69\cdot 10^{13}$\,cm$^{-2}$, and $\pm9.38\cdot 10^{13}$\,cm$^{-2}$, respectively. (g),(h) Doping-induced frequency shifts of the $E_{2g}$ and $A_{1g}$ phonons in MoS$_2$ as obtained with the pure nonadiabatic (NA, orchid squares) theory and with the NA theory corrected with the electron-hole scattering processes (NA+$\tau$, green squares). The experimental results from Ref.\,\cite{bib:ponomarev19} are shown with blue circles. (i),(j) Same as (g),(h) but for WS$_2$. (k) and (l) show the accompanying full-width-at-half-maximum (i.e., phonon linewidth) due the electron-hole pair scattering events for MoS$_2$ and WS$_2$, respectively.
}
\end{figure*}

{\bf Electron-hole pair scattering processes.} In what follows, I investigate the impact of the EPC-induced electron-hole pair (EHP) scattering processes\,\cite{bib:saitta08,bib:novko18} on phonons in TMDs. In Figs.\,\ref{fig:fig2}(a)-(d) the results of the phonon dispersions for different dopings and for WS$_2$ are shown along high-symmetry points of the Brillouin zone. Since the EHP rate is proportional to the EPC strength weighted with the first moment of the phonon spectrum, i.e., $1/\tau \propto \lambda\left\langle\omega\right\rangle$\,\cite{bib:novko18} (see the Methods section), the momentum- and mode-resolved EPC strengths weighted with frequencies $\lambda_{\mathbf{q}\nu}\omega_{\mathbf{q}\nu}$ are shown as well (orange circles). The doping induces changes in the EHP scattering rates in accordance with the transformations of the Fermi surface cuts. In particular, when solely K valleys intersect the Fermi level only the intra-valley EHP scatterings (i.e., $\mathrm{K}\rightarrow\mathrm{K}$ with $\mathbf{q}\approx\overline{\Gamma}$) are present. On the other hand, when multiple valleys are partially occupied or emptied both intra- and inter-valley EHP scatterings occur (e.g., $\mathrm{K}\leftrightarrow\Sigma$ with $\mathbf{q}\approx\overline{\rm M}$ for electron-doped and $\mathrm{K}\leftrightarrow\Gamma$ with $\mathbf{q}\approx\overline{\rm K}$ for hole-doped cases)\,\cite{bib:ge13,bib:piatti18}. In Figs.\,\ref{fig:fig2}(e)-(f) the results show how the EPC-induced EHP scattering rate as a function of doping depends greatly on the topology of the bands as well as on the SOC. For instance, the fingerprints of the Lifshitz transition are distinctly visible in the change of the $\lambda\left\langle\omega\right\rangle$ as a function of doping (e.g., through sudden increase for $0.08\,e/$unit cell in MoS$_2$ and for $-0.08\,e/$unit cell in WS$_2$). Furthermore, the SOC reduces the EHP scattering rates by more than factor of two in some cases. This is due to spin-orbit splitting of the valleys, where, e.g., the lower valence valley is more coupled to the phonons than the upper valence valley\,\cite{bib:hinsche17,bib:mahatha19} (see also the Supplementary Figure 8).

Finally, in Figs.\,\ref{fig:fig2}(g)-(h) and \ref{fig:fig2}(i)-(j), I show the effect of these EPC-induced EHP scatterings on the $E_{2g}$ and $A_{1g}$ phonon frequencies as a function of doping for MoS$_2$ and WS$_2$, respectively. The EHP scattering processes slightly washe out the NA corrections\,\cite{bib:saitta08,bib:novko18}, improving, for example, the agreement with the experiment in the case of WS$_2$ (see also the Supplementary Figure 5). However, a more striking and direct ramification of the EHP scattering events is the appearance of the phonon linewidth\,\cite{bib:saitta08,bib:novko18} [see Figs.\,\ref{fig:fig2}(k) and \ref{fig:fig2}(l)]. In close agreement with the experiments\,\cite{bib:chakraborty12,bib:lu17}, the linewidth of the $A_{1g}$ mode shows a steeper increase as a function of doping, and is generally larger than the $E_{2g}$ linewidth. This proves that the major part of the experimentally observed doping-induced phonon linewidth modifications in TMDs\,\cite{bib:chakraborty12,bib:lu17} is underlain by the EHP scattering processes due to the EPC (other contributions might be anharmonic coupling and scattering on inhomogeneities), as it is, for example, the case in MgB$_2$ where nonadiabatic effects as well play a significant role\,\cite{bib:novko18}. The linewidth, similarly to the frequency shift, shows a great sensitivity on the topology of the valleys, and therefore can also be used in Raman spectroscopy as an indicator of the Lifshitz transition.

\section*{Discussion}

All in all, the phonon dynamics in doped single-layer transition metal dichalcogenides visible in Raman spectra was shown to be quite complex and governed by the interplay between electron-electron, spin-orbit, and dynamical electron-phonon interactions. In particular, at the outset of the Lifshitz transition, i.e., when the Fermi level crosses multiple conduction or valence valleys, the adiabatic Born-Oppenheminer approximation breaks down and significant nonadiabatic effects are prompted. Further, the outset of the Lifshitz transition was shown to be dictated by the non-local electron-electron interaction, which determines the right topology of the valleys. In addition to that, the strong electron-hole pair scatterings due to electron-phonon coupling are present and responsible for the phonon linewidth increase as a function of doping. Note that these results are at odds with the previous studies, where the phonon spectrum of transition metal dichalcogenides was considered to be adiabatic\,\cite{bib:zhang15,bib:chakraborty12,bib:ponomarev19}.

\begin{figure}[!t]
\includegraphics[width=0.49\textwidth]{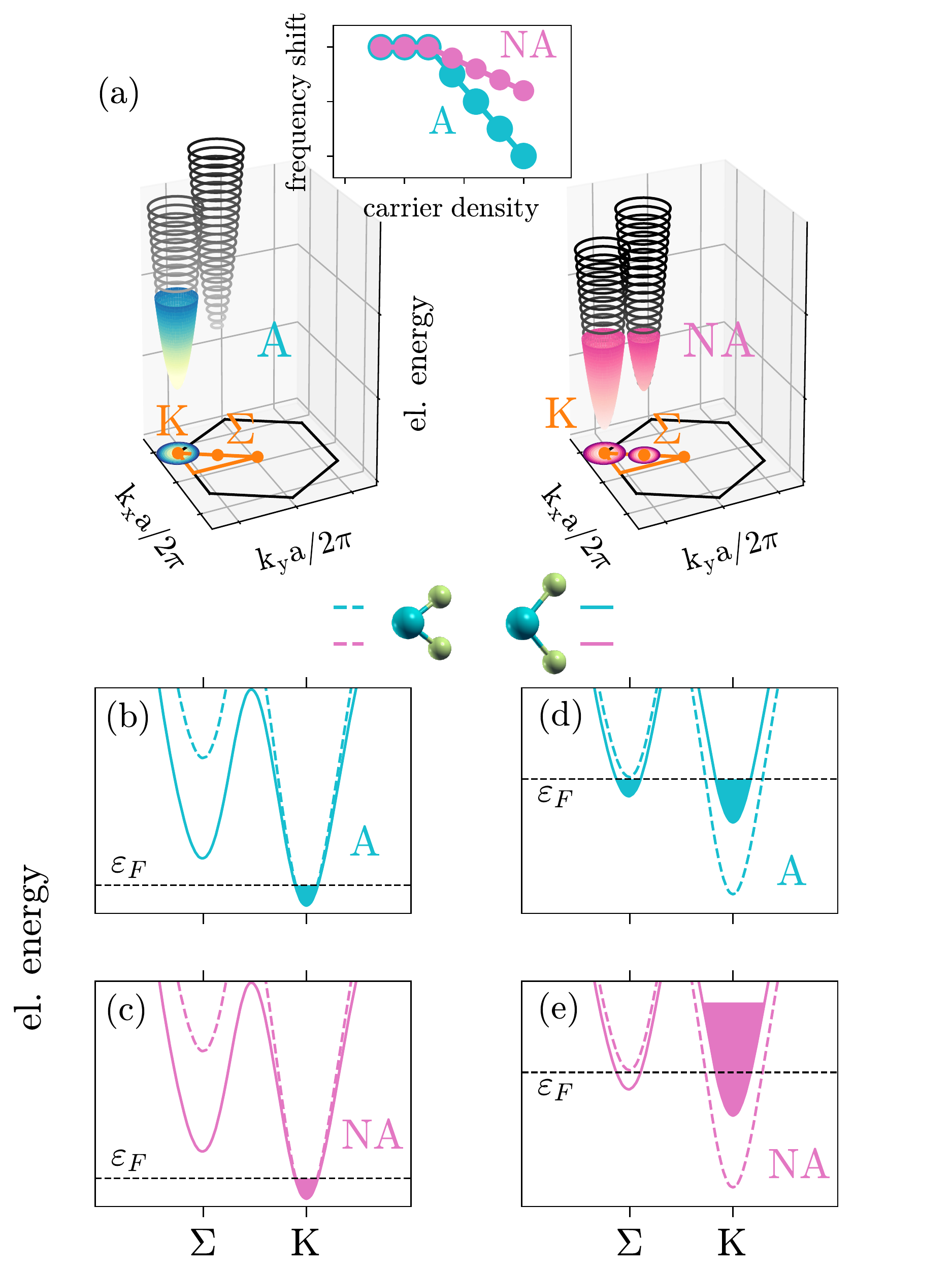}
\caption{\label{fig:fig3}{\bf Summary of multi-valley nonadiabatic effects in transition metal dichalcogenides.} (a) Schematic of electronic band structure of transition metal dichalcogenides in the case of the adiabatic (A) and nonadiabatic (NA) regimes are shown with light blue and orchid, respectively. Valleys are shown for each electron momentum along \textit{x} and \textit{y} directions, i.e., $k_x a / 2\pi$ and $k_y a / 2\pi$ (where \textit{a} is unit cell parameter). Inset: Schematic of electron-induced phonon frequency shifts as obtained with the A (light blue) and NA (orchid) approximations. In the A regime both A and NA approximations give zero frequency shifts, while in the NA regime these approximations produce disparate results. (b) A and (c) NA charge distribution modifications induced by the $A_{1g}$ phonon displacement when only the K valley is partially occupied (A regime). Dashed and full lines represent the bands when atoms are, respectively, at the equilibrium and displaced along the $A_{1g}$ eigenvectors. Black dashed horizontal line defines the Fermi energy $\varepsilon_F$. (d) A and (e) NA charge distribution modifications induced by the $A_{1g}$ phonon displacement when the K valley is partially occupied and the $\Sigma$ valley is almost crossing the Fermi level (NA regime). It turns out that static and dynamical charge modifications are similar and very different in the one-valley and multi-valley regimes, respectively. Analogous physical phenomena are expected in similar multi-valley semiconductors.
}
\end{figure}

The adiabatic and nonadiabatic regimes of single-layer transition metal dichalcogenides in the case of electron doping are schematically depicted in Fig.\,\ref{fig:fig3}(a). When only the K valley is partially occupied (the adiabatic regime), the modifications of the electronic structure induced by the $A_{1g}$ phonon does not generate the changes in the charge density, as shown in Figs.\,\ref{fig:fig3}(b) and \ref{fig:fig3}(c). Consequently, in this adiabatic regime, both adiabatic and nonadibatic approximations give the same results, i.e., zero frequency shifts [see the inset in Fig.\,\ref{fig:fig3}(a)]. On the other hand, at the outset of the Lifshitz transition (nonadiabatic regime), when the bottom of the $\Sigma$ valley is only slightly occupied, the atom displacements along the $A_{1g}$ mode give rise to charge modifications only for the adiabatic approximation [cf. Figs.\,\ref{fig:fig3}(d) and \ref{fig:fig3}(e)]. In that case, part of the charge from the K valley is distributed into the $\Sigma$ valley, where the electron-phonon coupling is much stronger (see the Supplementary Figure 8). For the nonadiabatic approximation, carriers remain in the same valley. As a result, the corresponding phonon frequency shifts turn out to be large for the adiabatic, while small for the nonadiabatic approximation.
It is important to stress that such multi-valley nonadiabatic mechanism was not discussed thus far\,\cite{bib:lazzeri06,bib:pisana07,bib:ferrante18,bib:saitta08,bib:quilty02,bib:caruso17}.

The insights given here might help elucidate the superconductive pairing mechanism in transition metal dichalcogenides, as well as puzzling discrepancy between the theoretically- and experimentally-obtained transition temperatures $T_c$. Even more, it is possible that the nonadiabatic channels investigated here could be universal and important in other novel multi-valley two-dimensional metallic (e.g., Nb$_2$C\,\cite{bib:huang19}) and semiconducting (e.g., black phosphorus\,\cite{bib:chakraborty16}) materials in the presence of the external excess charge.

\section*{Methods}

\textbf{Phonon self-energy.} Using the many-body perturbation theory, the phonon spectral function can be obtained with the following formula\,\cite{bib:novko18,bib:giustino17,bib:caruso17},
\begin{eqnarray}
B_{\nu}(\mathbf{q},\omega)=-\frac{1}{\pi}{\rm Im}\left[\frac{2\omega_{\mathbf{q}\nu}^{\mathrm{A}}}{\omega^2-(\omega_{\mathbf{q}\nu}^{\mathrm{A}})^2-2\omega_{\mathbf{q}\nu}^{\mathrm{A}}\widetilde{\pi}_{\nu}(\mathbf{q},\omega)}\right],
\label{eq:eq1}
\end{eqnarray}
where $\mathbf{q}$ and $\nu$ are the phonon momentum and band index, respectively, $\omega_{\mathbf{q}\nu}^{\mathrm{A}}$ is the adiabatic phonon frequency, and $\widetilde{\pi}_{\nu}$ is the nonadiabatic phonon self-energy due to the EPC (see the Supplementary Note 1). The real part of $\widetilde{\pi}_{\nu}$ gives the renormalization of the phonon frequency due to the NA coupling, i.e., $\omega^2=(\omega_{\mathbf{q}\nu}^{\mathrm{A}})^2+2\omega_{\mathbf{q}\nu}^{\mathrm{A}}\mathrm{Re}\,\widetilde{\pi}_{\nu}(\mathbf{q},\omega)$, while the imaginary part is the NA phonon linewidth, i.e.,  $\gamma_{\mathbf{q}\nu}=-2\mathrm{Im}\,\widetilde{\pi}_{\nu}(\mathbf{q},\omega)$. For simulating the Raman spectra only the $\mathbf{q}\approx0$ limits of $\widetilde{\pi}_{\nu}$ and $B_{\nu}$ are relevant. 

In the absence of the electron-hole pair scatterings the dynamical correction over the adiabatic phonon spectral function comes from the dynamical bare intraband phonon self-energy\,\cite{bib:lazzeri06,bib:saitta08,bib:giustino17,bib:novko18},
\begin{eqnarray}
\widetilde{\pi}_{\nu}^{\mathrm{0,intra}}(\omega)=\sum_{\mu\mathbf{k}} g^{b}_{\mu\mu,\nu}(\mathbf{k},0)g^{\ast}_{\mu\mu,\nu}(\mathbf{k},0) \left[-\frac{\partial f(\varepsilon_{\mu\mathbf{k}})}{\partial\varepsilon_{\mu\mathbf{k}}}\right].
\label{eq:eq3}
\end{eqnarray}
The electron band index and momentum are $\mu$ and $\mathbf{k}$, respectively, $\varepsilon_{\mu\mathbf{k}}$ is the corresponding electron energy, $f(\varepsilon_{\mu\mathbf{k}})$ is the Fermi-Dirac distribution function, while $g^{b}_{\mu\mu,\nu}$ and $g_{\mu\mu,\nu}$ are the bare and screened intraband electron-phonon coupling function. The dynamical interband contribution at $\mathbf{q}\approx 0$ is negligible for the TMDs, since the interband gap is much larger than the relevant phonon frequencies (see also the Supplementary Table 2). Equation\,\ref{eq:eq3} only contributes to the renormalization of the phonon frequency.
Note that the correct treatment of the Coulomb screening in the derivation of the phonon self-energy is employed here, i.e., with one screened and one bare electron-phonon coupling function\,\cite{bib:giustino17}, while the usual assumption is that both vertex functions are screened, i.e., $g_{\mu\mu',\nu}^{*}g_{\mu\mu',\nu}^{b}\rightarrow\left| g_{\mu\mu',\nu} \right|^2$. The results of Eq.\,\eqref{eq:eq3} for WS$_2$ and different dopings obtained with $g_{\mu\mu',\nu}^{*}g_{\mu\mu',\nu}^{b}$ and $\left| g_{\mu\mu',\nu} \right|^2$ are shown in Supplementary Figure 9. The corresponding difference is small, as it was already assumed in previous studies\,\cite{bib:saitta08,bib:calandra10}.

When EPC-induced electron-hole pair scattering processes up to all orders are taken into account\,\cite{bib:novko18,bib:cappelluti06,bib:saitta08}, the intraband phonon self-energy acquires the following form\,\cite{bib:novko18},
\begin{eqnarray}
\widetilde{\pi}_{\nu}^{\mathrm{intra}}(\omega)&=&\sum_{\mu\mathbf{k}} g^{b}_{\mu\mu,\nu}(\mathbf{k},0)g^{\ast}_{\mu\mu,\nu}(\mathbf{k},0)\left[-\frac{\partial f(\varepsilon_{\mu\mathbf{k}})}{\partial\varepsilon_{\mu\mathbf{k}}}\right]\nonumber\\
&&\times\frac{\omega}{\omega\left[1+\lambda_{\rm ep}(\omega)\right]+i/\tau_{\rm ep}(\omega)}.
\label{eq:eq4}
\end{eqnarray}
The scattering parameter describes the damping rate of the excited electron-hole pairs and for the case of electron-phonon scattering can be written as\,\cite{bib:novko18,bib:shulga91},
\begin{eqnarray}
1/\tau_{\rm ep}(\omega)
&=&
\frac{\pi}{\omega}
\int d\Omega \alpha^2 F(\Omega)
\left[
2\omega\coth\frac{\Omega}{2k_BT}\right.\nonumber\\
&&-
(\omega+\Omega)\coth\frac{\omega+\Omega}{2k_BT}\nonumber\\
&&\left.+
(\omega-\Omega)\coth\frac{\omega-\Omega}{2k_BT}
\right].
\label{eq:eq5}
\end{eqnarray}
where $k_B$ is the Boltzmann constant and $\alpha^2 F(\Omega)$ is the Eliashberg function. The dynamical energy renormalization parameter $\lambda_{\rm ep}(\omega)$ is obtained by performing the Kramers-Kronig transformation of $1/\tau_{\rm ep}(\omega)$. For $T=0$\,K and for frequencies much higher than the characteristic frequencies of the system, the damping rate acquires the following form $1/\tau_{\rm ep}(\infty)=\pi\sum_{\mathbf{q}\nu}\lambda_{\mathbf{q}\nu}\omega_{\mathbf{q}\nu}=\pi\lambda\left\langle\omega\right\rangle$, where $\lambda$ is the EPC constant and $\left\langle\omega\right\rangle$ is the so-called first moment of the phonon spectrum\,\cite{bib:novko17}.

\textbf{Computational details.} The ground-state calculations were done by means of the {\sc quantum espresso} (QE) package\,\cite{bib:qe1} with a plane-wave cutoff energy of 80 Ry. Fully relativistic norm-conserving pseudopotentials from the PseudoDojo project\,\cite{bib:pseudodojo} were used with the PBE exchange-correlation functional\,\cite{bib:pbe}. Spin-orbit coupling was also included. A ($24\times24\times1$) Monkhorst-Pack grid was used for sampling the Brillouin zone (with Gaussian smearing of 0.02\,Ry). Electron and hole dopings were simulated by adding and removing, respectively, the electrons and introducing the compensating homogeneous charged background\,\cite{bib:ge13}. 

In order to investigate the effects of the non-local electron-electron interactions on the electronic structure of the TMDs, the hybrid functionals PBE0\,\cite{bib:pbe0} and vdW-DF-cx0\,\cite{bib:cx0} were employed within QE. In addition, the corresponding electronic structures of MoS$_2$ and WS$_2$ were calculated within the self-consistent GW0\,\cite{bib:gw0} (i.e., where eigenvalues in the electronic Green's function are included self-consistently) implemented in {\sc gpaw}\,\cite{bib:gpaw}. Moderate differences between non-self-consistent G0W0 and self-consistent GW0 results are shown in the Supplementary Figure 1. The {\sc gpaw} calculations were done using the PAW pseudopotentials with energy cutoff of 600\,eV, PBE functional, and a ($18\times18\times1$) electron-momentum grid. Fermi-Dirac smearing was used with $T=300$\,K. The GW0 calculations were done on the highest valence and lowest conduction bands with the 100\,eV plane wave cutoff. The corresponding results of the valence and conduction band topology are in Table\,\ref{tab:table1}. Note that the comparison between different \emph{xc} approximations is made without spin-orbit coupling in Supplementary Figures 1-3. This is because the calculation with spin-orbit coupling along with hybrid DF-cx0 functional is not implemented in QE (the DF-cx0+SOC values presented in Table\,\ref{tab:table1} are actually the DF-cx0 values corrected with the energy modifications due to spin-orbit coupling as obtained with the PBE). Also, the band structure modifications induced by spin-orbit coupling are very similar at the GW0 and PBE levels, and are in agreement with ARPES experiments. Therefore, it is justified to make the fitting procedure described below without spin-orbit coupling, and then include it a posteriori (at the PBE level).

Since the phonon calculations are unfeasible with the hybrid functionals as well as with the GW approaximation, the further calculations were done with the PBE functional and with the exact amount of unit cell strain\,\cite{bib:yuan16,bib:ortenzi18} required to reproduce the same topology of the valleys as obtained with the hybrid vdW-DF-cx0 functional (the conclusions of the present paper would not change at all if the reference topology of valleys is as obtained with GW0 or PBE0). In particular, in the case of MoS$_2$ the unit cell strain of $\epsilon=-1.59\%$ ($a=3.135$\,\AA) and $\epsilon=0\%$ ($a=3.1856$\,\AA) were used, respectively, to simulate the right topology of the valence and conduction bands (which are then used for the hole and electron doping, respectively). For WS$_2$ the unit cell strain of $\epsilon=-1.64\%$ ($a=3.14$\,\AA) and $\epsilon=-0.54\%$ ($a=3.175$\,\AA) were used, respectively, to simulate the right topology of the valence and conduction bands. The distances between the Mo/W and S atoms were relaxed for each doping until forces were less than $10^{-5}$\,Ry/a.u. The spin-orbit coupling was then introduced for these strained structures and in the following phonon calculations. For further information see the Supplementary Figures 1-4.

Adiabatic phonon frequencies at the $\Gamma$ point were calculated on a dense ($72\times72\times1$) grid using density functional perturbation theory\,\cite{bib:baroni01} as implemented in QE and with the Gaussian smearing corresponding to $T=300$\,K. The intraband nonadiabatic phonon self-energies Eq.\,\eqref{eq:eq3} were extracted from the DFPT calculations as described in Supplementary Note 1 (the corresponding method is based on work presented in Y. Nomura et al.\,\cite{bib:nomura15}). The Eliashberg functions needed for calculating the relaxation rate Eq.\,\eqref{eq:eq5} were calculated with the {\sc epw} code\,\cite{bib:epw}, while the corresponding input quantities (electron energies, phonon frequencies, and electron-phonon coupling elements) were interpolated using maximally localized Wannier functions\,\cite{bib:wannier}.  The corresponding summations were done on $(480\times480\times1)$ electron and $(48\times48\times1)$ phonon momentum grids. The temperature of the Fermi-Dirac distribution in Eqs.\,\eqref{eq:eq3} and \eqref{eq:eq4} as well as the temperature entering Eq.\,\eqref{eq:eq5} was set to 300\,K as in the experiment\,\cite{bib:ponomarev19}. The phonon calculations were performed without and with 2D Coulomb interaction truncation, which eliminates the spurious interaction between periodic images (i.e., layers)\,\cite{bib:sohier16,bib:sohier17}. However, no difference between the two types of calculations were obtained (see the Supplementary Figure 10).



\section*{Data Availability}
The data that support the findings of this study are available from the author on reasonable request.

\section*{Code Availability}
The codes used to produce the data presented in this study are available from the author on reasonable request.

\bibliography{phtmd}

\section*{Acknowledgments}
The author gratefully acknowledges financial support from the European Regional Development Fund for the ``Center of Excellence for Advanced Materials and Sensing Devices'' (Grant No. KK.01.1.1.01.0001). Financial support by Donostia International Physics Center (DIPC) during various stages of this work is also highly acknowledged. Computational resources were provided by the DIPC computing center.

\section*{Author contributions}
D.N. conceived the idea, developed the method, performed theoretical calculations and analysis, created visualizations, and wrote the manuscript.

\section*{Competing interests}
The author declare no competing interests.

\section*{Additional information}
{\bf Supplementary information} is available for this paper at [URL].

%
%

%
%

%
%

\begin{table*}[hbt!]
\caption{\label{tab:table1}{\bf Topology details of valence and conduction valleys.} Energy differences between the top of the K and $\Gamma$ valleys in the valence band $\Delta\varepsilon^{v}_{{\rm K}\Gamma}$ and between the bottom of the $\Sigma$ and K valleys in the conduction band $\Delta\varepsilon^{c}_{\Sigma{\rm K}}$ for MoS$_2$ and WS$_2$ as obtained by means of several electron exchange and correlation approximations without and with spin-orbit coupling (SOC). In addition, the experimental results obtained in Refs.\,\citep{bib:miwa15,bib:bussolotti19,bib:eknapakul14,bib:dendzik15,bib:kastl18} are shown. All the values are given in meV.}
\begin{ruledtabular}
\begin{tabular}{llccccccccc}
 & & \multicolumn{8}{c}{\emph{xc} approximation} &  \\
\cline{3-10}
 & & PBE & PBE+SOC & GW0 & GW0+SOC & PBE0 & PBE0+SOC & DF-cx0 & DF-cx0+SOC & exp.
 \\ \hline
MoS$_2$ & $\Delta\varepsilon^{v}_{{\rm K}\Gamma}$ & $-$19 & 51  & 194 & 264 & 92  & 171 & 203 & 273 &  140-310\footnote{Refs.\,\cite{bib:miwa15,bib:bussolotti19}}  \\
& $\Delta\varepsilon^{c}_{\Sigma{\rm K}}$         &  275  & 244 & 54  & 23  & 431 & 427 & 265 & 234 &  $\gtrsim 60$\footnote{Ref.\,\cite{bib:eknapakul14}}  \\ \hline
WS$_2$& $\Delta\varepsilon^{v}_{{\rm K}\Gamma}$   &  46   & 231 & 164 & 349 & 103 & 356 & 232 & 417 &  240-510\footnote{Refs.\,\cite{bib:dendzik15,bib:kastl18}}   \\
& $\Delta\varepsilon^{c}_{\Sigma{\rm K}}$         &  190  & 84  & 112 & 6   & 338 & 209 & 137 & 31  &  --  \\
\end{tabular}
\end{ruledtabular}
\end{table*}

\end{document}